\documentclass[jmp,amsmath,amssymb,reprint]{revtex4-2}
\usepackage[pdftitle={Article},pdfauthor={Author},bookmarks=true,colorlinks,linkcolor=blue,urlcolor=blue,citecolor=blue,plainpages=false,pdfpagelabels,final,breaklinks=true]{hyperref}
\usepackage{orcidlink}
\usepackage{breakurl}
\usepackage{graphicx}
\usepackage{bm}
\usepackage[utf8]{inputenc}
\usepackage[T1]{fontenc}
\usepackage{mathptmx,mathrsfs,mathtools}
\usepackage{etoolbox,empheq}
\usepackage{enumitem}

\DeclareRobustCommand{\bmk}[1]{\bm{#1}}
\setlist[itemize,enumerate,description]{leftmargin=*}

\newcommand{\f}[2]{\frac{#1}{#2}}
\newcommand{\ko}[1]{\left( #1 \right)}
\newcommand{\kko}[1]{\left[ #1 \right]}

\newcommand{\tbx}[1]{\text{\small $ #1 $}}
\newcommand{\q}[1]{`#1'}
\newcommand{\dd}{\mathop{}\!d}

\newcommand{\rn}[1]{\uppercase\expandafter{\romannumeral #1\relax}}
\newcommand{\sv}[2]{#1\,|\,#2}
\newcommand{\dv}[3]{#1\,\|_{#3}\,#2}
\newcommand{\dvv}[3]{#1\,\|_{#3}\,#2}
\newcommand{\dvvv}[3]{#1\,\bigg\|_{#3}\,#2}

\DeclarePairedDelimiter\floor{\lfloor}{\rfloor}
\def\no{\nonumber}
\def\P{\bm{P}}
\def\Q{\bm{Q}}

\begin{document}

\title{On the $\bmk{q}$-generalised multinomial/divergence correspondence}
\author{Keisuke Okamura\,\orcidlink{0000-0002-0988-6392}}
\email[Corresponding author:~]{okamura@alumni.lse.ac.uk}
\affiliation{Embassy of Japan in the United States of America\\ \mbox{\footnotesize 2520 Massachusetts Avenue, N.W.\ Washington, DC 20008, USA.}}

\date{March 3, 2025}

\begin{abstract}
The asymptotic correspondence between the probability mass function of the $q$-deformed multinomial distribution and the $q$-generalised Kullback--Leibler divergence, also known as Tsallis relative entropy, is established.
The probability mass function is generalised using the $q$-deformed algebra developed within the framework of nonextensive statistics, leading to the emergence of a family of divergence measures in the asymptotic limit as the system size increases.
The coefficients in the asymptotic expansion yield Tsallis relative entropy as the leading-order term when $q$ is interpreted as an entropic parameter. 
Furthermore, higher-order expansion coefficients naturally introduce new divergence measures, extending Tsallis relative entropy through a one-parameter generalisation.
Some fundamental properties of these extended divergences are also explored.
\end{abstract}

\maketitle

\section{Introduction\label{sec:intro}}

A multinomial coefficient is a fundamental combinatorial quantity that features in a wide range of scientific fields, from pure and applied mathematics to information science and statistical physics.
Given a set of nonnegative integers $\bm{k}=\{k_{i}\}_{i=1}^{r}$ and $n=\sum_{i=1}^{r}k_{i}$, the multinomial coefficient is defined as
\begin{equation}\label{def:multinomial}
W_{1}(\bm{k})\coloneqq
\f{n!}{\prod_{i=1}^{r}k_{i}!}\,.
\end{equation}
This quantity represents the number of possible permutations of $n$ items across $r$ distinct categories, with $k_{i}$ items in category $i$.
It is well-known that, asymptotically, the logarithm of $W_{1}(\bm{k})$ yields Boltzmann--Gibbs--Shannon (BGS) entropy \cite{Gibbs02,Shannon48} as the leading term in the expansion for $n\sim k_{i}\gg 1$.
Specifically, by applying Stirling's approximation $\ln(n!)=n\ln n-n+O(n^{-1})$, similarly to $\ln(k_{i}!)$, we obtain the following result:
\begin{align}\label{asympt:cL1}
L_{1}(\P;n)\coloneqq\ln W_{1}(\P;n)
\,\sim\,
nH_{1}(\P)+O(\ln n)\,.
\end{align}
Here, $\P=\{p_{i}\}_{i=1}^{r}\equiv\{k_{i}/n\}_{i=1}^{r}$ represents the discrete probability distribution corresponding to $\bm{k}$, with $p_{i}\in [0,1]$ and $\sum_{i=1}^{r}p_{i}=1$.
The function $H_{1}(\P)\coloneqq -\sum_{i=1}^{r}p_{i}\ln p_{i}$ denotes the BGS entropy of distribution $\P$.
This combinatorial derivation of BGS entropy is known as Wallis's derivation \cite[p.~351]{Jaynes03}.

This Wallis's approach has recently been generalised to the $q$-deformed case for the complete order of the expansion in $n$ \cite{Okamura24}, within the framework of Tsallis's nonextensive statistics \cite{Tsallis88,Curado91,Abe01,TsallisBibList}.
The motivation for this $q$-generalisation arises from various physical phenomena and technical applications, such as memory effects, long-range interactions, and non-ergodic systems, which cannot be adequately described by BGS entropy and the corresponding additive statistics.
The resulting expression (see Eq.~(\ref{formula:cL}) later), which is the $q$-generalisation of Eq.~(\ref{asympt:cL1}), is represented as an infinite series expansion in terms of $n$, and involves Tsallis entropies \cite{Havrda67,Daroczy70,Tsallis88}---nonadditive $q$-generalisations of BGS entropy---with sequential entropic indices.
The formula gives an exact, smooth function for all real values of $q$, thus providing a distinctive characterisation of Tsallis entropy within $q$-deformed combinatorics.

In this paper, we extend our investigation to a generalisation of another classic result concerning the asymptotic relationship between the probability mass function (PMF) of a multinomial distribution and statistical divergence.
Divergence, also known as relative entropy or discrimination information \cite{Kullback51}, is a fundamental concept in statistics.
It serves as the basis for various important information-theoretic quantities, including joint entropy, cross-entropy, conditional entropy and mutual entropy \cite{Cover05}.
Divergence has applications across a broad spectrum of disciplines, such as statistical tests \cite{Hoeffding65,Arizono89}, model selection and inference \cite{Burnham02}, information geometry \cite{Amari16} and financial mathematics \cite{Kelly56}.
It also plays a significant role in pattern recognition \cite{Bai21}, with machine learning being a prominent modern application.

In the context of combinatorics, it is well known that Kullback--Leibler (KL) divergence \cite{Kullback51} appears as the leading term in the asymptotic expansion of the PMF of a multinomial distribution.
KL-divergence is a widely used measure for comparing two probability distributions.
Let $\Q=\{\pi_{i}\}_{i=1}^{r}$ be another discrete probability distribution, where $\pi_{i}\in [0,1]$ and $\sum_{i=1}^{r}\pi_{i}=1$.
The KL-divergence of $\P$ with respect to $\Q$ is then given by
\begin{equation}\label{PMF1}
f_{1}(\sv{\P}{\Q};n)
=W_{1}(\bm{k})\prod_{i=1}^{r}\pi_{i}^{k_{i}}\,.
\end{equation}
In the asymptotic limit where $n\sim k_{i}\gg 1$, the function $M_{1}(\dv{\P}{\Q}{};n)\coloneqq\ln f_{1}(\dv{\P}{\Q}{};n)$ behaves as
\begin{align}\label{asympt:cM1}
M_{1}(\dv{\P}{\Q}{};n)\,\sim\,
-nD_{1}(\dv{\P}{\Q}{})+O(\ln n)\,,
\end{align}
where $D_{1}(\dv{\P}{\Q}{})=\sum_{i=1}^{r}p_{i}\ln(p_{i}/\pi_{i})$ represents the KL-divergence of $\P$ with respect to $\Q$.

In the following sections, this paper will explore how the asymptotic formula for the multinomial distribution PMF can be extended to the $q$-generalised case, similar to the generalisation discussed earlier for entropy measures.
This extension is by no means straightforward. 
Firstly, it needs to be clarified how the multinomial distribution PMF should be $q$-deformed for this purpose. 
Secondly, it is uncertain whether the $q$-generalised KL-divergence, also known as Tsallis relative entropy \cite{Borland98,Tsallis98,Furuichi04,Huang16,Leinster19}, can be derived from this $q$-deformed PMF.
Nevertheless, we will demonstrate that Tsallis relative entropy naturally emerges from the asymptotic expansion of a particular $q$-deformed multinomial distribution PMF.
In doing so, we also naturally introduce new divergence measures, extending Tsallis relative entropy through a one-parameter generalisation.

The rest of this paper is organised as follows. 
In Section \ref{sec:qAsympt}, we review the formula for the asymptotic correspondence between the $q$-deformed multinomial coefficient and a family of Tsallis entropies, as presented in \cite{Okamura24}.
This section also includes relevant definitions and properties of $q$-operations and $q$-combinatorics.
We then derive the exact formula and analyse the structure of the function at particular characteristic values, limits of $q$, and in special cases of $\Q$.
Section \ref{sec:prop_D} investigates the properties of a newly introduced generalised divergence measure, of which Tsallis relative entropy is a particular case.
Finally, Section \ref{sec:conclusion} provides a summary and concluding remarks.

\section{The generalised multinomial/divergence correspondence\label{sec:qAsympt}}

\subsection{Review and preliminaries\label{subsec:preliminaries}}

The asymptotic formula for the $q$-generalised equivalent of the asymptotic correspondence in Eq.~(\ref{asympt:cL1}) has recently been derived in \cite{Okamura24}, which is valid for the entire range of real $q$.
This formula expresses the $q$-logarithm of the $q$-deformed multinomial coefficient \cite{Okamura24,Oikonomou07,Suyari06} as an infinite series expansion with respect to the system size $n$, combined with Tsallis entropies:
\begin{align}\label{formula:cL}
L_{q'}(\P;n)&\coloneqq\ln_{q'}W_{q'}(\P;n)\no\\
&\hspace{-2em}\,\sim\,\f{\zeta(1-q)}{1-q}H_{0}(\P)+\sum_{\ell\in\mathbb{Z}_{\geq 0}}^{}C_{\ell}(q)n^{q-\ell}H_{q-\ell}(\P)\,.
\end{align}
Here, the remainder terms, which tend to zero in the asymptotic limit, are omitted for clarity.
Below is an explanation of each term on the left-hand side (LHS) and the right-hand side (RHS) of the formula.

On the LHS of Eq.~(\ref{formula:cL}), the $q$-deformed multinomial coefficient \cite{Okamura24,Oikonomou07,Suyari06} is defined as
\begin{equation}\label{def:q-multinomial}
W_{q}(\bm{k})\coloneqq
n!_{q}\oslash_{q}\kko{\bigotimes_{i=1}^{n}{\!}_{q}\,k_{i}!_{q}}\,,
\end{equation}
where the $q$-logarithm \cite{Tsallis94} is defined by $\ln_{q}x\coloneqq\f{x^{1-q}-1}{1-q}$, which reduces to the ordinary logarithmic function $\ln x$ as $q\to 1$.
We adopt the convention that $0\ln_{q}(\cdot)=0$.
The $q$-multiplication ($\otimes_{q}$) and $q$-division ($\oslash_{q}$) are defined for two positive numbers $x$ and $y$, with $q\in\mathbb{R}{\setminus}\{1\}$, as follows \cite{Nivanen03,Borges04}:
\begin{equation}\label{def:q_multidiv}
x\,\Big\{{}^{\displaystyle\otimes_{q}}_{\displaystyle\oslash_{q}}\Big\}\,y
\coloneqq
\big[x^{1-q}\pm y^{1-q}\mp 1\big]_{+}^{\f{1}{1-q}}\,.
\end{equation}
The upper signs on the RHS correspond to $q$-multiplication, and the lower signs correspond to $q$-division, and $[\,\cdot\,]_{+}\coloneqq \max\{\,\cdot\,,0\}$.
The $q$-factorial ($!_{q}$) is defined for a nonnegative integer $n$ by $n!_{q}=\bigotimes_{\!q}{}_{\ell=1}^{n}{\!}\,\ell$.
In the limit as $q\to 1$, these operations reduce to their ordinary counterparts: $x\otimes_{q}y\to xy$, $x\oslash_{q}y\to x/y$, and $n!_{q}\to n!$.
Consequently, the $q$-multinomial coefficient (\ref{def:q-multinomial}) reduces to the ordinary multinomial coefficient (\ref{def:multinomial}).
In the LHS of Eq.~(\ref{formula:cL}), the $q$-logarithm of the $q$-multinomial coefficient is expressed using the \q{additive-dualised} entropic parameter $q'\equiv 2-q$ \cite{Curado91,Naudts04}, instead of $q$, for notational convenience.
For further insights into recent developments in $q$-deformed algebra and calculus, see \cite{Borges22,Borges22b}.

Next, on the RHS of Eq.~(\ref{formula:cL}), the function $H_{q}(\P)$ denotes the Tsallis entropy of distribution $\P$ with entropic parameter $q\in\mathbb{R}{\setminus}\{1\}$, defined as follows \cite{Havrda67,Daroczy70,Tsallis88}:
\begin{equation}\label{def:Tsallis_ent}
H_{q}(\P)\coloneqq
\f{1}{1-q}\kko{\sum_{i=1}^{r}(p_{i})^{q}-1}
=\sum_{i=1}^{r}p_{i}\ln_{q}\f{1}{p_{i}}\,,
\end{equation}
which recovers BGS entropy in the limit as $q\to 1$.
Tsallis entropy is a prominent example of nonadditive entropy that generalises the ordinary additive BGS entropy and plays a central role in Tsallis's nonextensive formalism \cite{Tsallis88,Curado91,Abe01}.
It is strictly concave when $q>0$ and strictly convex when $q<0$.

The function $\zeta(s)$ represents the Riemann zeta function, defined on $\mathbb{C}{\setminus}\{1\}$ as
\begin{equation}\label{def1:zeta}
\zeta(s)\coloneqq
\sum_{\ell=1}^{\infty}\f{1}{\ell^{s}}
\end{equation}
for $\mathfrak{R}(s)>1$, with analytic continuation elsewhere.
Finally, the coefficients $\{C_{\ell}(q)\}$ are given by
\begin{equation}\label{coeff}
C_{\ell}(q)=\binom{\ell-q+1}{\ell}\f{B_{\ell}}{q-\ell}\,,
\end{equation}
where $B_{\ell}$ represents the $\ell$-th Bernoulli number, with $B_{1}=-\f{1}{2}$ and $B_{\ell}=0$ for odd $\ell\geq 3$.

The behaviour of $L_{q}$ across characteristic values of $q$ was investigated in \cite{Okamura24}, with a summary provided in Appendix \ref{app:behav:cL}.
Wallis's classic result given by Eq.~(\ref{asympt:cL1}), which relates the multinomial coefficient to entropy, is reproduced as the $q\to 1$ limit of Eq.~(\ref{formula:cL}).

\subsection{The $\bmk{M}_{q}$ formula and the emergent divergence family\label{subsec:qDivergence}}

Having examined the multinomial coefficient, we now turn to the $q$-deformation of the full structure of the multinomial distribution PMF.
First, note that the definition in Eq.~(\ref{PMF1}) can equivalently be expressed as
\begin{equation}\label{PMF2}
f_{1}(\sv{\P}{\Q};n)
=\left.\prod_{\ell=1}^{n}\ell\right/\prod_{i=1}^{r}\prod_{\ell_{i}=1}^{k_{i}}\f{\ell_{i}}{\pi_{i}}\,.
\end{equation}
Guided by this expression, we define the $q$-deformed multinomial distribution PMF as
\begin{align}\label{def:probmass}
f_{q}(\sv{\P}{\Q};n)
\coloneqq\ko{\bigotimes_{\ell=1}^{n}{\!}_{q}\,\ell}\oslash_{q}\kko{\bigotimes_{i=1}^{r}{\!}_{q}\,\bigotimes_{\ell_{i}=1}^{k_{i}}{\!}_{q}\,\f{\ell_{i}}{\pi_{i}}}\,,
\end{align}
where the ordinary multiplication and division operations, except those between $\ell_{i}$ and $\pi_{i}$, are replaced by the $q$-multiplications and $q$-division operations defined in Eq.~(\ref{def:q_multidiv}).
Here, it is crucial to use ordinary division operation between $\ell_{i}$ and $\pi_{i}$ for each $i$ in the \q{denominator} (i.e.~the term to the right of $\oslash_{q}$).
This definition ensures that the $q$-deformed PMF yields an expression for $M_{q}(\dv{\P}{\Q}{};n)\coloneqq\ln_{q}f_{q}(\sv{\P}{\Q}{};n)$ that includes not only Tsallis relative entropy but also its extended variants.
Appendix \ref{app:alt} explores alternative definitions of the $q$-deformed multinomial distribution PMF and their implications.

As in Eq.~(\ref{formula:cL}), we use the dualised parameter $q'\equiv 2-q$ to take the $q$-logarithm for notational convenience.
Then, straightforward calculation yields:
\begin{align}
M_{q'}(\sv{\P}{\Q};n)
&=\sum_{\ell=1}^{n}\ln_{q'}\ell-\sum_{i=1}^{r}\sum_{\ell_{i}=1}^{k_{i}}\ln_{q'}\f{\ell_{i}}{\pi_{i}}\no\\
&=\f{1}{q-1}\kko{\sum_{\ell=1}^{n}\ell^{q-1}-\sum_{i=1}^{r}\sum_{\ell_{i}=1}^{k_{i}}\ko{\f{\ell_{i}}{\pi_{i}}}^{q-1}}\,.
\end{align}
In the limit as $n\to\infty$, the sum $\sum_{\ell=1}^{n}\ell^{q-1}$ converges to the Riemann zeta function $\zeta(1-q)$ if $q<0$.
To obtain an asymptotic formula for $M_{q'}(\dv{\P}{\Q}{};n)$ across the entire range of real $q$, we use the analytic continuation of the Riemann zeta function, as employed in \cite{Okamura24}.
This involves the analytically-continued expression of the Riemann zeta function defined by \cite[see e.g.][\href{http://dlmf.nist.gov/25.2.E9}{(25.2.9)}]{DLMF}
\begin{equation}\label{def:zeta}
\zeta(1-q)
=\sum_{\ell=1}^{n}\f{1}{\ell^{1-q}}+\sum_{\ell=0}^{m}\binom{\ell-q}{\ell}\f{B_{\ell}}{\ell-q}n^{q-\ell}-R_{n,m}(1-q)\,,
\end{equation}
which is valid for $q<m+1$ with $m\in\mathbb{Z}_{\geq 0}$.
Here, $R_{n,m}(1-q)$ represents the remainder term given by
\begin{equation}
R_{n,m}(1-q)=\binom{m+1-q}{m+1}\int_{n}^{\infty}\f{B_{m+1}(x-\floor*{x})}{x^{m+2-q}}\dd x\,,
\end{equation}
where $B_{\ell}(\cdot)$ denotes the $\ell$-th Bernoulli polynomial and $\floor*{x}$ represents the floor function, giving the largest integer less than or equal to $x$.
Using the analytically-continued expression, we have:
\begin{align}\label{cM:temp}
M_{q'}(\sv{\P}{\Q};n)&=\f{\zeta(1-q)}{q-1}\kko{1-\sum_{i=1}^{r}\pi_{i}^{1-q}}-\no\\
&\hspace{-3em}{}-\f{1}{q-1}\sum_{\ell=0}^{m}\binom{\ell-q}{\ell}\f{B_{\ell}}{\ell-q}\kko{n^{q-\ell}-\sum_{i=1}^{r}\pi_{i}^{1-q}k_{i}^{q-\ell}}+{}\no\\
&\hspace{-3em}{}+\f{1}{q-1}\kko{R_{n,m}(1-q)-\sum_{i=1}^{r}\pi_{i}^{1-q}R_{k_{i},m}(1-q)}\,.
\end{align}

To proceed, we introduce the following quantity defined between the two distributions $\P$ and $\Q$ with entropic parameter $q\in\mathbb{R}{\setminus}\{1\}$ and real parameter $\lambda$:
\begin{align}\label{def:cD}
D_{q}(\dv{\P}{\Q}{\lambda})
&\coloneqq\f{1}{1-q}\kko{1-\sum_{i=1}^{r}p_{i}^{q}\pi_{i}^{1-q-\lambda}}\no\\
&=-\sum_{i=1}^{r}p_{i}\ln_{q}\Big(\pi_{i}^{\f{1-q-\lambda}{1-q}}\Big/p_{i}\Big)\,.
\end{align}
The case $\lambda=0$ corresponds to Tsallis relative entropy \cite{Borland98,Tsallis98,Furuichi04,Huang16,Leinster19}, which is defined by
\begin{equation}\label{def:q-KL}
D_{q}(\dv{\P}{\Q}{0})
=-\sum_{i=1}^{r}p_{i}\ln_{q}\f{\pi_{i}}{p_{i}}\,.
\end{equation}
Its limit as $q$ approaches $1$ reproduces the ordinary KL-divergence \cite{Kullback51}: $\lim_{q\to 1}D_{q}(\dv{\P}{\Q}{0})=D_{1}(\dv{\P}{\Q}{})$.
The current paper refers to the extended divergence measure defined in Eq.~(\ref{def:cD}) as the $\lambda$-extended Tsallis relative entropy.
Its fundamental properties will be detailed later in Section \ref{sec:prop_D}.

In terms of the $\lambda$-extended Tsallis relative entropy, the second block of terms in Eq.~(\ref{cM:temp}) can be expressed as:
\begin{align}
&-\f{1}{q-1}\sum_{\ell=0}^{m}\binom{\ell-q}{\ell}\f{B_{\ell}}{\ell-q}n^{q-\ell}\kko{1-\sum_{i=1}^{r}\pi_{i}^{1-q}p_{i}^{q-\ell}}\no\\
&\hspace{2em}=-\sum_{\ell=0}^{m}C_{q}(\ell)n^{q-\ell}D_{q-\ell}(\dv{\P}{\Q}{\ell})\,,
\end{align}
where the coefficients $\{C_{\ell}(q)\}$ are the same as given in Eq.~(\ref{coeff}).
Note also that the first block of terms in Eq.~(\ref{cM:temp}) is proportional to $D_{0}(\dv{\P}{\Q}{q})$.
Finally, to ensure that the above expression is valid for the entire $q\in\mathbb{R}{\setminus}\{1\}$, we take the limit as $m\to\infty$.
For sufficiently large $n\to\infty$, the remainder terms tend to zero, and Eq.~(\ref{cM:temp}) behaves as:
\begin{align}\label{formula:cM}
M_{q'}(\sv{\P}{\Q};n)
&\,\sim\,\f{\zeta(1-q)}{q-1}D_{0}(\dv{\P}{\Q}{q})-{}\no\\
&\hspace{2em}{}-\sum_{\ell\in\mathbb{Z}_{\geq 0}}C_{\ell}(q)n^{q-\ell}D_{q-\ell}(\dv{\P}{\Q}{\ell})\,.
\end{align}
This extends the correspondence between the multinomial distribution PMF and KL-divergence to the $q$-deformed scenario.
One can observe that formula (\ref{formula:cM}) can be formally derived from Eq.~(\ref{formula:cL}) related to the $q$-deformed multinomial coefficient by substituting $H_{q-\ell}(\P)$ with $-D_{q-\ell}(\dv{\P}{\Q}{\ell})$.

\subsection{Characteristic values and limits\label{subsec:characteristic}}

Let us investigate some characteristic values and limits of formula (\ref{formula:cM}).
The term involving the analytically continued Riemann zeta function is crucial for ensuring the smoothness of $M_{q}(\sv{\P}{\Q};n)$ over the entire range of $q\in\mathbb{R}$, as it eliminates the divergences arising from $C_{0}(q)$ and $C_{1}(q)$ when $q=2$ and $q=1$, respectively.
This property is shared with $L_{q}(\P;n)$; see \cite{Okamura24}.
By straightforward calculation, the functional forms of $M_{q}(\sv{\P}{\Q};n)$ at specific values of $q$ are:
\begin{subequations}
\begin{align}
{M}_{-2}&=-\f{n^{4}}{4}D_{4,0}-\f{n^{3}}{3}D_{3,1}-\f{n^{2}}{12}D_{2,2}\,,\label{cM:q->-2}\\
{M}_{-1}&=-\f{n^{3}}{3}D_{3,0}-\f{n^{2}}{4}D_{2,1}\,,\label{cM:q->-1}\\
{M}_{0}&=-\f{n^{2}}{2}D_{2,0}\,,\label{cM:q->0}\\
{M}_{1}&=-nD_{1,0}+\f{1}{2}\ln(2\pi n)D_{0,1}-\f{1}{2}\ln\prod_{i=1}^{r}p_{i}+\f{D_{-1,2}}{6n}-{}\no\\
&\quad{}-\f{D_{-3,4}}{90n^{3}}+\f{D_{-5,6}}{210n^{5}}-\f{D_{-7,8}}{210n^{7}}+\dots\,,\label{cM:q->1}\\
{M}_{2}&=\ln\prod_{i=1}^{r}p_{i}-\f{D_{-1,1}}{n}+\f{D_{-2,2}}{4n^{2}}-\f{D_{-4,4}}{24n^{4}}+\f{D_{-6,6}}{36n^{6}}+\dots\,.\label{cM:q->2}
\end{align}
\end{subequations}
Here, we use the shorthand notation $M_{q}\equiv M_{q}(\sv{\P}{\Q};n)$ and $D_{q,\ell}\equiv D_{q}(\dv{\P}{\Q}{\ell})$ for notational simplicity.
Note that the term $-nD_{1,0}$ in Eq.~(\ref{cM:q->1}) corresponds to $-n$ times the KL-divergence of $\P$ with respect to $\Q$.
In other words, Eq.~(\ref{cM:q->1}) completes the asymptotic expansion given in Eq.~(\ref{asympt:cM1}).

For $q>2$, the term involving the Riemann zeta function becomes the leading term in the expansion of $M_{q}$ with respect to $n$.
For sufficiently large $q\gg 2$, the asymptotic limit is given by
\begin{align}\label{cM:q>>2}
{M}_{q\gg 2}(\sv{\P}{\Q};n)&\,\sim\,-\f{\zeta(q-1)}{q-1}D_{0}(\dv{\P}{\Q}{q'})\no\\
&\,\sim\,-\f{1}{q}\bigg(1-\sum_{i=1}^{r}\pi_{i}^{q-1}\bigg)\,.
\end{align}
Subsequently, it is evident that $\lim_{q\to\infty}M_{q}=0$.
On the other extreme, when $q\to -\infty$, the function $M_{q}$ scales as $n^{-q}$.

\subsection{The $\bmk{P}\equiv\bmk{Q}$ case\label{subsec:P=Q}}

When $\P\equiv\Q$, the function $M_{q}$ simplifies and can be expressed in terms of a family of Tsallis entropies.
Specifically, using the dualised entropic parameter $q'\equiv 2-q$, we have:
\begin{align}
M_{q'}(\sv{\P}{\P};n)
&\sim\,\f{q\zeta(1-q)}{1-q}H_{1-q}(\P)+{}\no\\
&\hspace{-1em}{}+\f{1}{q-1}\sum_{\ell\in\mathbb{Z}_{\geq 0}}\binom{\ell-q-1}{\ell-1}B_{\ell}n^{q-\ell}H_{1-\ell}(\P)\,,
\end{align}
where we use the relations provided later in Eq.~(\ref{D:l:P=Q}).
The limit as $q$ (or equivalently $q'$) approaches 1 is:
\begin{align}\label{P=Q,q=1}
M_{1}(\sv{\P}{\P};n)&=\lim_{q\to 1}M_{q}(\sv{\P}{\P};n)\no\\
&\hspace{-5.2em}\,\sim\,-\f{1}{2}\ln(2\pi n)H_{0}-\f{1}{2}\ln\prod_{i=1}^{r}p_{i}-\f{H_{-1}}{6n}+\f{H_{-3}}{90n^{3}}+\dots\,,
\end{align}
and it follows that:
\begin{equation}
L_{1}(\P;n)=nH_{1}(\P)+M_{1}(\sv{\P}{\P};n)\,.
\end{equation}
Thus, BGS entropy times the system size, $n$, is exactly given by the difference between the multinomial coefficient and the multinomial distribution PMF, considering full-order accuracy in the expansion in terms of $n$.
This difference corresponds to the term of the form $\prod_{i=1}^{r}\pi_{i}^{k_{i}}$, where $\pi_{i}=p_{i}=nk_{i}$.

\section{Properties of the $\lambda$-extended Tsallis relative entropy\label{sec:prop_D}}

This section explores some fundamental properties of the newly introduced divergence measure, the $\lambda$-extended Tsallis relative entropy, with a particular emphasis on its similarities and differences compared to ordinary Tsallis relative entropy.
The trivial case of a single-category scenario (i.e.~$p_{i}=1$ for some $i$) is excluded from the discussion.

\subsection{The $\bmk{\lambda=0}$ case\label{subsec:l=0}}

The $\lambda=0$ case represents Tsallis relative entropy, given by Eq.~(\ref{def:q-KL}), which recovers the original KL-divergence in the limit as $q\to 1$.
Its properties, some of which we present below, have been investigated in the literature \cite{Borland98,Tsallis98,Furuichi04,Huang16,Leinster19}.
We do not limit the range of $q$ to $q>0$ (or $q\geq 0$), but consider the entire range of $q\in\mathbb{R}$.

When the two distributions are element-wise equivalent, i.e.~$p_{i}=\pi_{i}$ for all $i$, Tsallis relative entropy of $\P$ with respect to $\Q(\,\equiv\P)$ vanishes:
\begin{equation}\label{D:0:P=Q}
D_{q}(\dv{\P}{\P}{0})=0\,.
\end{equation}
Otherwise, this measure has a definite sign depending on $q$.
Specifically, if $\P\not\equiv\Q$, it satisfies:
\begin{description}
\item[Sign-definiteness]
\begin{equation}\label{D:0}
D_{q}(\dv{\P}{\Q}{0})
\gtreqless 0
~~\text{if}~~
q\gtreqless 0\,,
\end{equation}
where the inequality and equality signs at the top, middle and bottom correspond to those in the conditions, respectively.
This property follows from Jensen's inequality based on the curvature-definiteness of the function $-\ln_{q}(\cdot)$, which is strictly convex if $q>0$ and strictly concave if $q<0$.

\end{description}
Tsallis relative entropy is known to satisfy additional properties, as listed below:
\begin{description}
\item[Continuity]
$D_{q}(\dv{\P}{\Q}{0})$ is continuous with respect to all elements of $\P=\{p_{i}\}_{i=1}^{r}$ and $\Q=\{\pi_{i}\}_{i=1}^{r}$.
\item[Expansibility]
$D_{q}(\dv{\P}{\Q}{0})$ is invariant if $p_{r+1}=0$ and $\pi_{r+1}=0$ are added to $\P$ and $\Q$, respectively.
Specifically, if we denote $\widetilde{\P}=\{p_{1},\,\dots,\,p_{r},0\}$ and $\widetilde{\Q}=\{\pi_{1},\,\dots,\,\pi_{r},0\}$,
\begin{equation}\label{Expansibility:0}
D_{q}(\dv{\P}{\Q}{0})=D_{q}(\dv{\widetilde{\P}}{\widetilde{\Q}}{0})\,.
\end{equation}
\item[Symmetry]
Let $\P_{\sigma}=\{p_{\sigma(i)}\}_{i=1}^{r}$ and $\Q_{\sigma}=\{\pi_{\sigma(i)}\}_{i=1}^{r}$ denote permutations $\sigma$ of $\P$ and $\Q$, respectively.
Then
\begin{equation}\label{Symmetry:0}
D_{q}(\dv{\P}{\Q}{0})=D_{q}(\dv{\P_{\sigma}}{\Q_{\sigma}}{0})\,.
\end{equation}
\item[Pseudoadditivity]
For independent probability distributions $\P^{(a)}=\big\{p_{i}^{(a)}\big\}_{i=1}^{r_{a}}$ and $\Q^{(a)}=\big\{\pi_{i}^{(a)}\big\}_{i=1}^{r_{a}}$, where $a=1,\,2$, and which satisfy $\sum_{i=1}^{r_{a}}p_{i}^{(a)}=\sum_{i=1}^{r_{a}}\pi_{i}^{(a)}=1$,
\begin{align}\label{Pseudoadditivity:0}
&D_{q}\big(\dvv{\P^{(1)}\otimes\P^{(2)}}{\Q^{(1)}\otimes\Q^{(2)}}{0}\big)\no\\
&=D_{q}\big(\dvv{\P^{(1)}}{\Q^{(1)}}{0}\big)\oplus_{q'}D_{q}\big(\dvv{\P^{(2)}}{\Q^{(2)}}{0}\big)\,,
\end{align}
where $\P^{(1)}\otimes\P^{(2)}\coloneqq\big\{p_{i}^{(1)}p_{j}^{(2)}\,\big|\,0\leq i\leq r_{1},\,0\leq j\leq r_{2}\big\}$ and similar for $\Q^{(1)}\otimes\Q^{(2)}$.
The $q$-addition \cite{Nivanen03,Borges04} is given by $x\oplus_{q}y\coloneqq x+y+(1-q)xy$.
This property can be proven by direct calculation.
\item[Recursivity]
Let $p_{\star}=p_{1}+p_{2}$, $\pi_{\star}=\pi_{1}+\pi_{2}$, $\P_{\star}=\{p_{\star},\,p_{3},\,\dots,\,p_{r}\}$ and $\Q_{\star}=\{\pi_{\star},\,\pi_{3},\,\dots,\,\pi_{r}\}$.
Then
\begin{align}\label{Recursivity:0}
D_{q}(\dv{\P}{\Q}{0})
&=D_{q}(\dv{\P_{\star}}{\Q_{\star}}{0})+{}\no\\
&\hspace{-1em}{}+p_{\star}^{q}\pi_{\star}^{1-q}D_{q}\tbx{\bigg(\dvvv{\bigg\{\f{p_{1}}{p_{\star}},\,\f{p_{2}}{p_{\star}}\bigg\}}{\bigg\{\f{\pi_{1}}{\pi_{\star}},\f{\pi_{2}}{\pi_{\star}}\bigg\}}{0}\bigg)}\,.
\end{align}
This relation can be proven by direct calculation.
\item[Joint curvature-definiteness]
For independent probability distributions $\P^{(a)}=\big\{p_{i}^{(a)}\big\}_{i=1}^{r}$ and $\Q^{(a)}=\big\{\pi_{i}^{(a)}\big\}_{i=1}^{r}$, where $a=1,\,2$, and which satisfy $\sum_{i=1}^{r}p_{i}^{(a)}=\sum_{i=1}^{r}\pi_{i}^{(a)}=1$, the following holds for $0\leq\alpha\leq 1$:
\begin{align}\label{curvature:0}
&D_{q}\big(\dv{\P_{\alpha}}{\Q_{\alpha}}{0}\big)
\lesseqgtr
\alpha D_{q}\big(\dvv{\P^{(1)}}{\Q^{(1)}}{0}\big)+{}\no\\
&\hspace{6em}{}+(1-\alpha)D_{q}\big(\dvv{\P^{(2)}}{\Q^{(2)}}{0}\big)
~~\text{if}~~
q\gtreqless 0\,,
\end{align}
where $\P_{\alpha}=\alpha\P^{(1)}+(1-\alpha)\P^{(2)}$ and similar for $\Q$.
In other words, the function $D_{q}(\dv{\P}{\Q}{0})$ is strictly convex about its both arguments for $q>0$, and is strictly concave about its both arguments for $q<0$.
Equality holds when $\alpha=0$ or $1$, as well as when $q=0$.
\end{description}

\subsection{The $\bmk{\lambda\neq 0}$ case\label{subsec:l}}

Next, we examine the nonzero $\lambda$ case.
When $\P\equiv\Q$, the $\lambda$-extended Tsallis relative entropy, defined for $q\in\mathbb{R}{\setminus}\{1\}$, is given by
\begin{equation}\label{D:l:P=Q}
D_{q}(\dv{\P}{\P}{\lambda})
=\f{\lambda}{q-1}H_{1-\lambda}(\P)\,.
\end{equation}
Thus, unlike the $\lambda=0$ case in Eq.~(\ref{D:0:P=Q}), this divergence does not vanish even when $\P\equiv\Q$, but is proportional to $\lambda$.
It is important to note that the $q\to 1$ limit for this expression alone is not well-defined.
However, when combined with the first term in Eq.~(\ref{formula:cM}), which involves the Riemann zeta function, it yields a well-defined finite value; see Eq.~(\ref{P=Q,q=1}).

The relationship (\ref{D:l:P=Q}) also implies that $D_{q}(\dv{\P}{\P}{\lambda})$ is nonnegative if $\lambda(q-1)>0$, and nonpositive if $\lambda(q-1)<0$.
However, when $\P\not\equiv\Q$, this relation needs to be adjusted.
Specifically, the $\lambda$-extended Tsallis relative entropy satisfies:
\begin{description}
\item[Sign-definiteness]
For $\lambda\geq 1$,
\begin{equation}\label{D:l}
D_{q}(\dv{\P}{\Q}{\lambda})
\gtrless 0
~~\text{if}~~
q\gtrless 1\,.
\end{equation}
Notably, the sign of $D_{q}(\dv{\P}{\Q}{\lambda})$ no longer depends on the sign of $q$, but rather on the sign of $q-1$.
This contrasts with the case of Tsallis relative entropy ($\lambda=0$), as seen in Ineq.~(\ref{D:0}).

This property can be proven as follows.
First, we use the fact that the function $-\ln_{q}(\cdot)$ is strictly convex if $q>0$ and strictly concave if $q<0$.
Applying Jensen's inequality yields
\begin{align}\label{ineq1}
D_{q}(\dv{\P}{\Q}{\lambda})
&\gtreqless -\ln_{q}\ko{\sum_{i=1}^{r}p_{i}\cdot\pi_{i}^{\f{1-q-\lambda}{1-q}}\bigg/p_{i}}\no\\
&\hspace{-3em}=-\ln_{q}\sum_{i=1}^{r}\pi_{i}^{\f{1-q-\lambda}{1-q}}
~~\text{if}~~
q\gtreqless 0~~(\text{and}~q\neq 1)\,.
\end{align}
Equality holds if and only if $\pi_{i}=\f{1}{r}$ for all $i$, or if $q=0$.
In addition, since the function $-\ln_{q}(\cdot)$ is monotonically decreasing on $(0,\infty)$ for all $q$, 
\begin{align}\label{ineq2}
-\ln_{q}\sum_{i=1}^{r}\pi_{i}^{\f{1-q-\lambda}{1-q}}
\gtrless -\ln_{q}\sum_{i=1}^{r}\pi_{i}
=0
~~\text{if}~~
\lambda(q-1)\gtrless 0\,,
\end{align}
where we used the fact that $\pi_{i}^{\f{1-q-\lambda}{1-q}}=\pi_{i}\cdot\pi_{i}^{\f{\lambda}{q-1}}\gtrless \pi_{i}$ for each $i$ if $\lambda(q-1)\lessgtr 0$.
From Ineqs.~(\ref{ineq1},\,\ref{ineq2}), it can be deduced that $D_{q}(\dv{\P}{\Q}{\lambda})$ is positive if $\lambda>0$ and $q>1$, or if $\lambda<0$ and $0\leq q<1$, while it is negative if $\lambda>0$ and $q\leq 0$.
For other combinations of $q$ and $\lambda$ intervals, the sign of $D_{q}(\dv{\P}{\Q}{\lambda})$ is not straightforward and requires further case separation, or the sign remains undetermined depending on the interval.
However, when $\lambda\geq 1$ and $0<q<1$, it can be shown that $D_{q}(\dv{\P}{\Q}{\lambda})<0$ holds.
To show this, note that $\pi_{i}^{\f{1-q-\lambda}{1-q}}>1$ if $1-\lambda<q<1$.
Therefore, when $\lambda\geq 1$ and $0<q<1$, the inequality $\pi_{i}^{\f{1-q-\lambda}{1-q}}>1$ is satisfied.
Additionally, the function $-\ln_{q}(\cdot)$ is monotonically decreasing on $(0,\infty)$.
Using these properties, the $\lambda\,(\geq 1)$-extended Tsallis relative entropy for $0<q<1$ can be evaluated as
\begin{align}\label{ineq3}
D_{q}(\dv{\P}{\Q}{\lambda})
&=-\sum_{i=1}^{r}p_{i}\ln_{q}\Big(\pi_{i}^{\f{1-q-\lambda}{1-q}}\Big/p_{i}\Big)\no\\
&< -\sum_{i=1}^{r}p_{i}\ln_{q}\f{1}{p_{i}}
=-H_{q}(\P)\leq 0\,,
\end{align}
where the last inequality follows from the nonnegativity of Tsallis entropy (\ref{def:Tsallis_ent}).
Combining all these results above, the proposition (\ref{D:l}) is proven for $\lambda\geq 1$.
Note that Ineq.~(\ref{ineq3}) also holds if $\lambda<0$ and $1<q<1-\lambda$, or if $0<\lambda<1$ and $1-\lambda<q<1$, which provides another interval where $D_{q}(\dv{\P}{\Q}{\lambda})$ is negative-definite.
\item[Continuity, Expansibility, Symmetry and Pseudoadditivity]
The $\lambda\,(\neq 0)$-extended Tsallis relative entropy also satisfies these properties in the same form as for Tsallis relative entropy (when $\lambda=0$).
These properties can be obtained by replacing $\dv{}{}{0}$ in Eqs.~(\ref{Expansibility:0},\,\ref{Symmetry:0},\,\ref{Pseudoadditivity:0}) with $\dv{}{}{\lambda\,(\neq 0)}$.
We omit the detailed display here.
\item[Recursivity]
Defining the $\star$-variables in the same way as in Eq.~(\ref{Recursivity:0}), we have:
\begin{align}\label{Recursivity:l}
D_{q}(\dv{\P}{\Q}{\lambda})
&=D_{q}(\dv{\P_{\star}}{\Q_{\star}}{\lambda})+{}\no\\
&\hspace{-1em}{}+p_{\star}^{q}\pi_{\star}^{1-q-\lambda}D_{q}\tbx{\bigg(\dvvv{\bigg\{\f{p_{1}}{p_{\star}},\,\f{p_{2}}{p_{\star}}\bigg\}}{\bigg\{\f{\pi_{1}}{\pi_{\star}},\f{\pi_{2}}{\pi_{\star}}\bigg\}}{\lambda}\bigg)}\,.
\end{align}
Note that the exponent on $\pi_{\star}$ on the RHS is $1-q-\lambda$, in contrast to the Tsallis relative entropy case (when $\lambda=0$).
This property can be proven by direct calculation.
\item[Joint curvature-definiteness]
The $\lambda\,(\neq 0)$-extended Tsallis relative entropy does not generally satisfy the joint curvature-definiteness property when $\dv{}{}{0}$ in Eq.~(\ref{curvature:0}) is replaced by $\dv{}{}{\lambda\,(\neq 0)}$.
Counterexamples are easy to find.
Specifically, even when $q>0$, the inequality may or may not hold, depending on the choice of $\P$ and $\Q$; the same applies when $q\leq 0$.
\end{description}

\subsection{Relationship to Tsallis entropy\label{subsec:vs_Tsallis}}

Let $\Q_{\mathrm{uni}}=\{\pi_{i}=\f{1}{r}\}_{i=1}^{r}$ represent the uniform distribution.
In the KL-divergence case, there is a straightforward relationship with BGS entropy, given by $D_{1}(\dv{\P}{\Q_{\mathrm{uni}}}{0})=\ln r-H_{1}(\P)$.
This relationship can be extended to the $q$-generalised form, connecting the $\lambda$-extended Tsallis relative entropy and Tsallis entropy, as follows:
\begin{equation}
D_{q}(\dv{\P}{\Q_{\mathrm{uni}}}{\lambda})
=\tbx{\ko{\f{1-q-\lambda}{1-q}}}H_{2-q-\lambda}(\Q_{\mathrm{uni}})-r^{q-1}H_{q}(\P)\,.
\end{equation}
When $\lambda=0$, this simplifies to $D_{q}(\dv{\P}{\Q_{\mathrm{uni}}}{0})=H_{2-q}(\Q_{\mathrm{uni}})-r^{q-1}H_{q}(\P)=r^{q-1}\big(\ln_{q}r-H_{q}(\P)\big)$ \cite{Furuichi04}.
This relation implies, in light of Ineq.~(\ref{D:0}), that the degree-$q$ Tsallis entropy of $\P$ is bounded above by $\ln_{q}r$ for $q>0$, and bounded below by $\ln_{q}r$ when $q<0$.

\section{Summary and Conclusion\label{sec:conclusion}}

In this paper, we have investigated an asymptotic correspondence between the $q$-deformed multinomial distribution PMF, a fundamental $q$-deformed combinatorial quantity formulated herein, and the $q$-generalised KL-divergence, also known as Tsallis relative entropy \cite{Borland98,Tsallis98,Furuichi04,Huang16,Leinster19}.
First, the $q$-deformed multinomial distribution PMF was constructed by applying the $q$-deformed algebra \cite{Nivanen03,Borges04}, which is associated with Tsallis's nonextensive statistics \cite{Tsallis88,Curado91}, to the ordinary multinomial distribution PMF.
Special attention was given during this construction to ensure that the resulting $q$-deformed multinomial distribution PMF naturally leads to the $q$-generalised divergence measures, including Tsallis relative entropy.

The resulting expression gives rise to a family of divergence measures in the asymptotic limit, as shown in Eq.~(\ref{formula:cM}).
The asymptotic expansion is smooth for all real values of $q$ due to the analytic continuation of the Riemann zeta function, resulting from the $q$-deformation procedure.
The expansion coefficients yield Tsallis relative entropy as the leading-order term when $q$ is interpreted as an entropic parameter.
This result provides valuable insights from a combinatorial perspective, offering a deeper mathematical and physical understanding of the Tsallis statistical mechanics framework, which is renowned for its intriguing physical phenomena and applications in information science across various values of $q$.

Furthermore, higher-order expansion coefficients naturally introduce new divergence measures, as defined in Eq.~(\ref{def:cD}), referred to as the $\lambda$-extended Tsallis entropies in this paper.
This new measure extends Tsallis relative entropy through a one-parameter generalisation.
Its fundamental properties, such as sign-definiteness, pseudoadditivity and recursivity, are also investigated.
Generally, the more flexible a divergence measure, the greater its potential for diverse applications in information science.
Therefore, the $\lambda$-extended Tsallis relative entropy may provide specific applications depending on the context, extending beyond its connection to the $q$-deformed multinomial distribution PMF.

\begin{acknowledgments}
The views and conclusions contained herein are those of the author and should not be interpreted as necessarily representing the official policies or endorsements, either expressed or implied, of any of the organisations with which the author is currently or has been affiliated in the past.
\end{acknowledgments}

\section*{Data Availability Statement}
Data sharing is not applicable to this article as no new data were created or analysed in this study.

\newpage
\appendix


\section{Asymptotic behaviours of $\bmk{L}_{q}$\label{app:behav:cL}}

The functional forms of $L_{q}(\P;n)$ at specific values of $q$ are presented as follows \cite{Okamura24}:
\begin{subequations}
\begin{align}
{L}_{-2}&=\f{n^{4}}{4}H_{4}+\f{n^{3}}{3}H_{3}+\f{n^{2}}{12}H_{2}\,,\label{cL:q->-2}\\
{L}_{-1}&=\f{n^{3}}{3}H_{3}+\f{n^{2}}{4}H_{2}\,,\label{cL:q->-1}\\
{L}_{0}&=\f{n^{2}}{2}H_{2}\,,\label{cL:q->0}\\
{L}_{1}&=nH_{1}-\f{1}{2}\ln(2\pi n)H_{0}-\f{1}{2}\ln\prod_{i=1}^{r}p_{i}-\f{H_{-1}}{6n}+\f{H_{-3}}{90n^{3}}-{}\no\\
&\quad{}-\f{H_{-5}}{210n^{5}}+\f{H_{-7}}{210n^{7}}+\dots\,,\label{cL:q->1}\\
{L}_{2}&=\ln\prod_{i=1}^{r}p_{i}+\gamma H_{0}+\f{H_{-1}}{n}-\f{H_{-2}}{4n^{2}}+\f{H_{-4}}{24n^{4}}-\f{H_{-6}}{36n^{6}}+\dots\,,\label{cL:q->2}
\end{align}
where $\gamma$ in Eq.~(\ref{cL:q->2}) is Euler's constant and we use the shorthand notations $L_{q}\equiv L_{q}(\P;n)$ and $H_{q}\equiv H_{q}(\P)$.
\end{subequations}

\section{Note on the $\bmk{q}$-deformation of the multinomial distribution PMF\label{app:alt}}

In the main text, we defined the $q$-deformed multinomial distribution PMF as given in Eq.~(\ref{def:probmass}).
This appendix explores the results obtained if alternative definitions are considered.

For instance, suppose we adopt the following \q{na\"{i}ve} definition:
\begin{align}\label{def:PMF1a}
f_{q}^{\langle\text{A1}\rangle}(\sv{\P}{\Q};n)
&=\overbrace{\Bigg\{\bigg(\underbrace{\bigotimes_{\ell=1}^{n}{\!}_{q}\,\ell}_{\text{\footnotesize ${}=n!_{q}$}}\bigg)\oslash_{q}\bigg[\bigotimes_{i=1}^{r}{\!}_{q}\,\underbrace{\bigotimes_{\ell_{i}=1}^{k_{i}}{\!}_{q}\,\ell_{i}}_{\text{\footnotesize ${}=k_{i}!_{q}$}}\bigg]\Bigg\}}^{\text{\footnotesize $q$-deformed multinomial coefficient}}
\,\otimes_{q}\no\\
&\quad\otimes_{q}\overbrace{\bigg[\bigotimes_{i=1}^{r}{\!}_{q}\,\bigg(\pi_{i}^{\bigotimes\nolimits{\!}_{q}\,k_{i}}\bigg)\bigg]}^{\text{\footnotesize \q{$q$-def.} of $\prod_{i=1}^{r}\pi_{i}^{k_{i}}$}}\,.
\end{align}
On the RHS, inside the big curly bracket represents the $q$-deformed multinomial coefficient \cite{Okamura24,Oikonomou07,Suyari06}.
The other factor $q$-multiplied to it is the \q{(na\"{i}vely) $q$-deformed} $\prod_{i=1}^{r}\pi_{i}^{k_{i}}$.

Alternatively, another \q{na\"{i}ve} approach would suggest that the multinomial distribution PMF is $q$-deformed to be
\begin{align}\label{def:PMF1b}
\hspace{-0.65em}f_{q}^{\langle\text{A2}\rangle}(\sv{\P}{\Q};n)
=\ko{\bigotimes_{\ell=1}^{n}{\!}_{q}\,\ell}\oslash_{q}\kko{\bigotimes_{i=1}^{r}{\!}_{q}\,\bigotimes_{\ell_{i}=1}^{k_{i}}{\!}_{q}\,\big(\ell_{i}\oslash_{q}\pi_{i}\big)}\,,
\end{align}
where each $\ell_{i}/\pi_{i}$ in Eq.~(\ref{def:probmass}) has been replaced with $\ell_{i}\oslash_{q}\pi_{i}$, by using $q$-division.

Although it may initially appear that the two alternative definitions (\ref{def:PMF1a}) and (\ref{def:PMF1b}) differ, a direct calculation using $q$-deformed operations reveals that they are, in fact, identical.
By following the same procedure outlined in \cite{Okamura24} and in the current paper, it can be shown that both definitions lead to the following expression when taking the $q$-logarithm (with the dualised parameter $q'\equiv 2-q$ as used in the main text):
\begin{align}\label{M:PMF1}
\ln_{q'}f_{q'}^{\langle\text{A}\rangle}(\sv{\P}{\Q};n)
&=L_{q}(\P;n)-n\sum_{i=1}^{r}p_{i}\ln_{q}\f{1}{\pi_{i}}\,.
\end{align}
This expression is identical to the $q'$-logarithm of the $q'$-multinomial coefficient, as given in (\ref{formula:cL}), except that the additional term involving $-n\sum_{i=1}^{r}p_{i}\ln_{q}\f{1}{\pi_{i}}$ indicates the deviation from the $L_{q}(\P;n)$ formula.
As can be seen, there are no nontrivial divergence structures present in Eq.~(\ref{M:PMF1}).

Yet another way to $q$-deform the multinomial distribution PMF would be to replace each $\ell_{i}/\pi_{i}$ in Eq.~(\ref{def:probmass}) with $\ell_{i}\otimes_{q}\f{1}{\pi_{i}}$, by using $q$-multiplication:
\begin{align}\label{def:PMF2}
\hspace{-0.65em}f_{q}^{\langle\text{B}\rangle}(\sv{\P}{\Q};n)
=\ko{\bigotimes_{\ell=1}^{n}{\!}_{q}\,\ell}\oslash_{q}\kko{\bigotimes_{i=1}^{r}{\!}_{q}\,\bigotimes_{\ell_{i}=1}^{k_{i}}{\!}_{q}\,\bigg(\ell_{i}\otimes_{q}\f{1}{\pi_{i}}\bigg)}.
\end{align}
This yields, when taking the $q'$-logarithm,
\begin{align}\label{M:PMF2}
\ln_{q'}f_{q'}^{\langle\text{B}\rangle}(\sv{\P}{\Q};n)
&=L_{q}(\P;n)+n\sum_{i=1}^{r}p_{i}\ln_{q}\pi_{i}\,,
\end{align}
indicating, again, that there are no nontrivial divergence structures in Eq.~(\ref{M:PMF2}).
The deviation from the $L_{q}(\P;n)$ formula is $n\sum_{i=1}^{r}p_{i}\ln_{q}\pi_{i}$, which is related to the deviation term in (\ref{M:PMF1}) through the \q{additive duality} $q\leftrightarrow 2-q$.

In contrast, through the \q{canonical} formulation of the $q$-deformed multinomial distribution PMF adopted in Eq.~(\ref{def:probmass}), we verified in the main text that it indeed produces divergence measures, including Tsallis relative entropy and its $\lambda$-extended variants.


%

\end{document}